\documentclass[twocolumn,prl,preprintnumbers,tightenlines,nofootinbib,superscriptaddress]{revtex4}

\pdfoutput=1
\usepackage[english]{babel}
\usepackage{amsmath,amssymb,amsfonts}
\usepackage{graphicx}
\usepackage{booktabs}
\usepackage{hyperref}
\usepackage{color}
\usepackage[sort&compress]{natbib}
\usepackage{tikz}
\usepackage{cleveref}
\newcounter{qnumber}

\begin{document}

\preprint{IPMU19-0108}
\preprint{DESY 19-138}
\preprint{KEK-TH-2147}
\preprint{KEK-Cosmo-241}
\title{Testing Seesaw and Leptogenesis with Gravitational Waves}

\author{Jeff A. Dror}
\email{asafjeffdror@gmail.com}
\affiliation{Department of Physics, University of California, Berkeley, CA 94720, USA}
\affiliation{Ernest Orlando Lawrence Berkeley National Laboratory, Berkeley, CA 94720, USA}

\author{Takashi Hiramatsu}
\email{hiramatz@icrr.u-tokyo.ac.jp}
\affiliation{Institute for Cosmic Ray Research, The University of Tokyo, Kashiwanoha 5-1-5, Kashiwa 277-8582, Japan}

\author{Kazunori Kohri}
\email{kohri@post.kek.jp}
\affiliation{Institute of Particle and Nuclear Studies, KEK, 1-1 Oho, Tsukuba 305-0801, Japan}
\affiliation{The Graduate University for Advanced Studies (SOKENDAI), 1-1 Oho, Tsukuba 305-0801, Japan}
\affiliation{Kavli Institute for the Physics and Mathematics of the
  Universe (WPI), University of Tokyo,
  Kashiwa 277-8583, Japan}

\author{Hitoshi Murayama}
\email{hitoshi@berkeley.edu, Hamamatsu Professor}
\affiliation{Department of Physics, University of California, Berkeley, CA 94720, USA}
\affiliation{Kavli Institute for the Physics and Mathematics of the
  Universe (WPI), University of Tokyo,
  Kashiwa 277-8583, Japan}
\affiliation{Ernest Orlando Lawrence Berkeley National Laboratory, Berkeley, CA 94720, USA}
\affiliation{Deutsches Elektronen-Synchrotron DESY, Notkestrasse 85, 22607 Hamburg, Germany}

\author{Graham White}
\email{GWhite@triumf.ca}
\affiliation{TRIUMF, 4004 Wesbrook Mall, Vancouver, BC V6T 2A3, Canada}

\begin{abstract} 
We present the possibility that the seesaw mechanism with thermal leptogenesis can be tested using the stochastic gravitational background. Achieving neutrino masses consistent with atmospheric and solar neutrino data, while avoiding non-perturbative couplings, requires right handed neutrinos lighter than the typical scale of grand unification. This scale separation suggests a symmetry protecting the right handed neutrinos from getting a mass. Thermal leptogenesis would then require that such a symmetry be broken below the reheating temperature. We enumerate all such possible symmetries consistent with these minimal assumptions and their corresponding defects, finding that in many cases, gravitational waves from the network of cosmic strings should be detectable. Estimating the predicted gravitational wave background we find that future space-borne missions could probe the entire range relevant for thermal leptogenesis.
\end{abstract}

\maketitle

\section{Introduction}
The discovery of masses and mixings of neutrinos~\cite{Fukuda:1998mi} marked the first robust evidence for physics beyond the Standard Model (SM) of particle physics. Interestingly, the masses are much smaller compared to those of the other elementary matter particles. It has become a pressing question how to understand the finite yet tiny neutrino masses theoretically. 

Arguably, the most popular mechanism to explain the smallness of the neutrino masses is the so-called seesaw mechanism~\cite{Yanagida:1980xy,Minkowski:1977sc,GellMann:1980vs} as it explains two puzzles simultaneously: tiny neutrino masses and origin of the asymmetry between matter and anti-matter in the Universe.  In its simplest incarnation, the Type-I seesaw, new SM-singlet fermions (right-handed neutrinos, $ N $) are introduced whose masses are much higher than the electroweak scale - a natural possibility as they are not forbidden by any symmetry. If the right-handed neutrino mass ($ M _R $) is below the reheating temperature of the universe, they will quickly be produced after inflation. Right handed neutrinos are inherently unstable and their eventual decay to a Higgs and a lepton can pick up CP violation in the Yukawa couplings, resulting in a preferential decay into anti-leptons. Subsequently, the anomalous violation of baryon and lepton numbers in the Standard Model partially converts the negative lepton asymmetry to the positive baryon asymmetry.  This scenario is called thermal leptogenesis \cite{Fukugita:1986hr}. The existence of right-handed neutrinos is further natural when the Standard Model gauge groups are unified into an $SO(10)$ grand unified theory.  Here and below, whenever we refer to the seesaw mechanism, it is meant to be Type-I seesaw together with thermal leptogenesis.

Unfortunately, the seesaw mechanism is notoriously difficult to test experimentally.  For successful thermal leptogenesis, the right-handed neutrino mass must be above $ \gtrsim  10^{9}$~GeV (see, {\it e.g.}\/, \cite{Buchmuller:2004nz}), and cannot be tested by terrestrial experiments.\footnote{The scale of leptogenesis can be brought lower if the reheating temperature is below the seesaw scale \cite{HahnWoernle:2008pq} and lower again if there is a mass degeneracy \cite{Pilaftsis:2003gt} or a fine tuning \cite{Hambye:2003rt}. The scale of supersymmetric leptogenesis can also be lower \cite{Hamaguchi:2001gw,Grossman:2005yi}.}  Therefore conceivable tests of the seesaw mechanism rely on circumstantial evidence, such as neutrino-less double beta decay \cite{DellOro:2016tmg}, CP violation in neutrino oscillation \cite{Endoh:2002wm,Esteban:2016qun}, structure in the mixing matrix \cite{Bertuzzo:2010et}, or indirect constraints relying on vacuum meta-stability \cite{Ipek:2018sai,Croon:2019dfw}.  It is therefore highly desirable to find other evidence to test the neutrino sector.

For the seesaw mechanism to have at least one neutrino with mass $ m _\nu \gtrsim 0.1~ {\rm eV} $  and the Yukawa coupling remaining perturbative below the grand unification (GUT) scale,  the right handed neutrino masses cannot be arbitrarily large giving the rough bound: $M_{R} \lesssim  10^{15}$~GeV. This scale is parametrically lower than the Planck scale or a possible GUT scale (typically chosen to be $ V \sim 10 ^{ 16} ~{\rm GeV} $) and suggests a possible symmetry that forbids the mass of the right-handed neutrinos. Assuming there are no large mass hierarchies among the right handed neutrinos, leptogenesis requires the Hubble scale during inflation to be above this scale and hence predicts a phase transition. If this phase transition leads to formation of topological defects, we expect stochastic gravitational wave from dynamics of the defect network.

In this {\em Letter}, we point out that the stochastic gravitational waves from the cosmic string network is quite a generic prediction of the seesaw mechanism. We enumerate all possible symmetries that could protect the right handed neutrino mass and point out their predicted defect structure. A common possibility seen in different breaking structures is the persistence of a cosmic string network. We compute the gravitational wave spectrum and compare with projections from future space missions, finding that such experiments could probe most of the parameter space necessary for thermal leptogenesis.

\section{Symmetry Breaking Patterns}
We begin by showing that the cosmic string network is a generic prediction of the seesaw mechanism when $B-L$ is broken spontaneously, rather than explicitly. For this purpose, we classify all possible symmetry breaking patterns. 
 
We require that there is an extended gauge symmetry $G$ which forbids the mass for the right-handed neutrinos, is flavor-blind, and is broken below the Hubble scale during inflation to allow for leptogenesis.  As a minimalist approach, we consider gauge symmetries that are at most rank 5\footnote{With the standard model particle content with right-handed Majorana neutrinos, the only possible low-energy discrete gauge symmetries are Z2 matter parity we considered and Z3 baryon number, yet the latter is broken in most higher gauge theories.  Therefore, as long as the Z2 matter parity is a subgroup of higher gauge symmetries, the most likely consequence is the cosmic strings based on this Z2, no matter how high the rank of higher gauge symmetry is} and are non-anomalous with only the standard-model fermions and right-handed neutrinos (while not the focus of this work, we note that non-minimal gauge groups would offer additional opportunities to look for topological defects). We also require that the symmetry breaking from $G$ to the Standard Model gauge group, $G_{\rm SM} = [SU(3)_{C} \times SU(2)_{L} \times U(1)_{Y}]/{\mathbb Z}_{6}$, does not lead to magnetic monopoles, allowing the symmetry breaking to occur below the inflationary scale.  With these assumptions,we find that there is only a finite set of possible gauge groups:
\begin{align}
	G_{\rm disc} &= G_{\rm SM} \times {\mathbb Z}_{N} \,,\\
	G_{B-L} &= G_{\rm SM} \times U(1)_{B-L} \,,\\
	G_{LR} &= SU(3)_{C} \times SU(2)_{L} \times SU(2)_{R} \times U(1)_{B-L} \,,\\
	G_{421} &= SU(4)_{\rm PS} \times SU(2)_{L} \times U(1)_{Y} \,,\\
	G_{\rm flip} &= SU(5) \times U(1)\,.
\end{align}
For the first case, ${\mathbb Z}_{N}$ is a discrete subgroup of the $U(1)_{B-L}$ gauge group, and the right handed neutrino mass is forbidden for $N \geq 3$.  For instance, it could be the ${\mathbb Z}_{4}$ center of $SO(10)$. $ G _{ B - L } $ is the extension of the SM to $ B - L $ which forbids the right handed neutrino mass as they carry lepton number, and $ U(1) _{ B - L } $ plays a similar role in $ G _{ LR} $. $SU(4)_{\rm PS}$ unifies $SU(3)_{C}$ and $U(1)_{B-L}$ in a way that originally appeared in the Pati--Salam theory, $G_{\rm PS} = SU(4)_{\rm PS} \times SU(2)_{L} \times SU(2)_{R}$~\cite{Pati:1974yy}, where now the right handed neutrino mass term would transform under the $ SU(4) _{ \rm PS} $. The last case is often called flipped SU(5)~\cite{Barr:1981qv} and here the right handed neutrinos are charged under the new $ U(1) $. Note that all of the above can be embedded into a unified $SO(10)$ gauge group.

On the other hand, one can also ask the question whether there can be a discrete gauge group below the mass scale of right-handed neutrinos.  By requiring that the discrete gauge group is non-anomalous under $SU(3)_{C}$, $SU(2)_{L}$, and gravity, one can show that the only possibility is the matter parity ${\mathbb Z}_{2}$ that flips the signs of all quarks and leptons but nothing else. Namely, the symmetry breaking pattern is either $G \rightarrow H=G_{\rm SM}$ or $G \rightarrow H=G_{\rm SM} \times {\mathbb Z}_{2}$.  Whether the matter parity remains unbroken depends on the representation of the Higgs field that generates the mass of the right-handed neutrinos.\footnote{Note that the matter parity can be identified with the ${\mathbb Z}_{2}$ subgroup of the ${\mathbb Z}_{4}$ center of $SO(10)$.  This is reminiscent of the $SO(10)$ origin of the $R$-parity in the Minimal Supersymmetric Standard Model.}

When $G$ is further embedded into larger groups such as $SO(10)$, topological defects may be unstable.  For instance, when $G_{N}$ is embedded into a connected group such as $SO(10)$ or $G_{B-L}$, the domain wall is unstable against the spontaneous creation of a string loop via quantum tunneling. There, the string loop grows to destroy the entire wall. Similarly, when $G_{B-L}$ is embedded into a simply-connected group such as $SO(10)$ or $G_{\rm PS}$, the string is unstable due to the spontaneous pair-creation of a monopole and an anti-monopole. This cuts the string, which shrinks and disappears. We explore these effects further below. 

We now study the stochastic gravitational wave background predicted by breaking patterns which induce cosmic strings. The gravitational wave spectrum has been studied in~\cite{Buchmuller:2013lra} as a consequence of $G_{B-L}$, including hybrid inflation based on the same gauge group as well as supersymmetry, in particular the gravitino problem.  As we noted here, the cosmic string network is far more general.  On the other hand, the consequences of inflation and supersymmetry are more model-dependent, and we focus on the symmetry breaking alone.

\section{Gravitational Waves from Strings}
The stochastic gravitational wave prediction from a cosmic string network has been highly controversial. A conventional estimate relies on Nambu--Goto string, an approximation where the string is infinitely thin with no couplings to particles \cite{Vachaspati:1984gt}. In this case, the numerical simulations are tractable over a large range of distance scales and hence frequencies of gravitational waves. There is additional uncertainty in the loop length ($ l _i $) at the time of formation ($ t _i $) which is normally taken to be a linear relation: $ l _i = \alpha t _i $. The parameter $ \alpha $ has a peaked distribution in both radiation and matter domination ranging from $ 0.01 - 0.1 $~\cite{Blanco-Pillado:2013qja}. 

Unfortunately, there has been major disagreements whether the particle production dominates the energy loss over that from gravitational wave emission.  Simulations based on Nambu--Goto strings cannot address this question.  If particle production dominates~\cite{Hindmarsh:2011qj}, the resulting stochastic gravitational wave background is suppressed by the {\it quadratic} power in $G\mu$~\cite{Buchmuller:2013lra} (where $ G $ is Newton's constant and $\mu $ is the string tension and roughly given by the square of the symmetry breaking scale, $ \mu \sim v ^2  $). Recent work in~\cite{Matsunami:2019fss} did extensive numerical simulations with the abelian Higgs model and found that the particle production is only important for extremely small loops, and hence the gravitational wave is the dominant mechanism for most situations. The present study is only for the BPS string (the critical point where the gauge boson mass is equal to the Higgs mass of the symmetry breaking scalar) but we suspect there is no qualitative change for non-BPS strings, as both the Higgs and gauge bosons are massive. On the other hand, the gravitational wave emission may be further enhanced if the difference between the gravitational radiation scale and gravitational back reaction scale is considered (see, {\it e.g.}\/, \cite{Ringeval:2017eww}). This possibility is under active study \cite{Blanco-Pillado:2019nto}. We assume the dominance of the gravitational wave emission in this paper, but emphasize that the discrepancy among various estimates needs to be settled before concrete predictions can be made.
To estimate the gravitational wave emission we follow the strategy employed in~\cite{Cui:2018rwi} which assumes large loops are produced with a spectrum sharply peaked at a given $ \alpha $, which we fix to be $ 0.05 $, and a fraction of energy released in the form of GW of $ {\cal F} _\alpha \simeq 0.1 $. The energy density ($ \Omega _{ {\rm GW}} $) per unit $ \log f $ (where $ f $ is the frequency) can be derived for each string normal-mode, $ k $ (see~\cite{Cui:2018rwi} for more details),
\begin{align} 
\Omega _{ {\rm GW}} & = \sum _{k = 1 } ^{\infty}  \Omega _{ {\rm GW}} ^{ (k) } ( f ) \,,\\ 
\Omega _{ {\rm GW}} ^{ (k) } & = \Omega _0 ^{(k)} ( f ) \int _{ 1 } ^{ \tau  _0 } d \tau \frac{ C _{ {\rm eff}} ( \tau  _i ) }{ \tau  _i ^4 } \frac{ a ^2 (\tau ) a ^3  ( \tau  _i ) }{ a _0 ^5 } \Theta ( \tau  _i  - \tau  _F ) \,,\\ 
\Omega _0 ^{(k)} ( f ) & = \frac{1}{ \rho _c } \frac{ 2 k }{ f } \frac{ {\cal F} _\alpha \Gamma ^{(k)} G \mu ^2 }{ \alpha ^2 t _F ^3 }  \,,\\ 
\tau _i ( \tau ) & = \frac{1}{ \alpha } \left[ \frac{ 2 k }{ f t _F } \frac{ a (\tau ) }{ a _0 } + \Gamma G \mu \tau  \right] \,,
\end{align} 
where $ \tau _a \equiv  t _a / t _F $,  $ t _F $ is the time the cosmic string network reaches the scaling regime (shortly after symmetry breaking), $ C _{ \rm eff} = 0.5 $ ($ 5.7 $) in matter (radiation) domination, $ \Gamma ^{(k)}  \simeq  \Gamma k ^{ - 4/3} / 3.6$ is a dimensionless constant which parameterizes the emission rate per mode, $ \Gamma \simeq 50 $, $ \Theta  $ is the Heaviside theta function which restricts string production till after formation of the scaling regime, $ a $ is the scale factor, and $ \rho _c $ is the critical density. 

\begin{figure} 
  \begin{center} 
    \includegraphics[width=8.5cm]{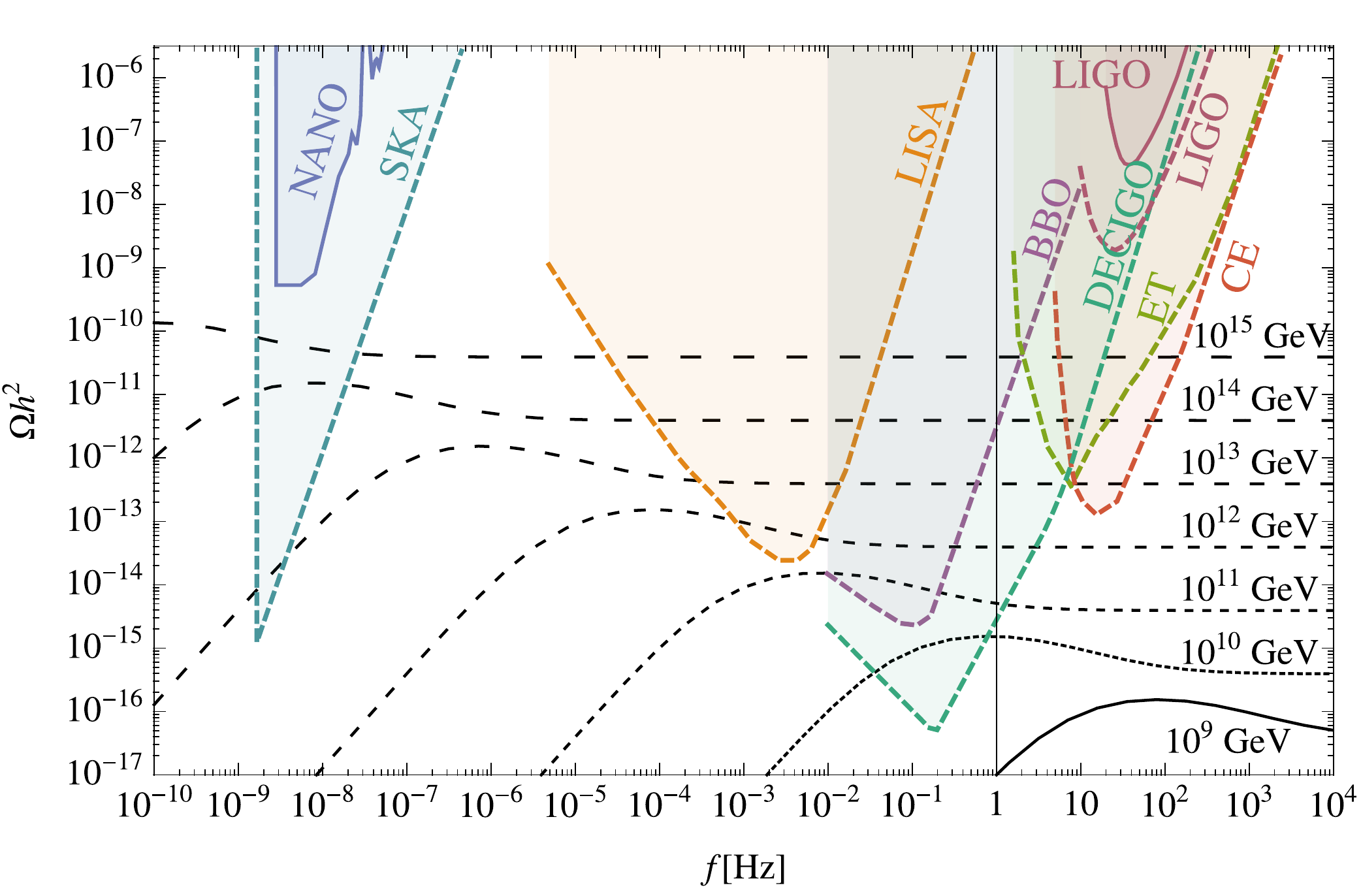} 
\end{center}
\caption{The predicted GW background from cosmic strings for different symmetry breaking scales, assuming the particle production is subdominant. For comparison we also display the sensitivity of current (solid) and future (dashed) experiments (from left to right) of Square Kilometer Array (SKA), NANOGRAV (NANO), Laser Interferometer Space Antenna (LISA), Big Bang Observer (BBO), DECi-hertz Interferometer Gravitational wave Observatory (DECIGO), Einstein Telescope (ET), Cosmic Explorer (CE), and Laser Interferometer Gravitational-Wave Observatory (LIGO). Here, we made an approximation for the string tension $\mu = v^{2}$ where $v$ is the symmetry breaking scale.}
\label{fig:stringGW}
\end{figure}

We present the stochastic gravitational wave background for different symmetry breaking scales assuming a simple radiation domination to matter domination cosmology in Fig.~\ref{fig:stringGW}. The flat scale invariant contribution arises from radiation domination and remains all the way up to frequencies beyond expected future capabilities. The additional bump at lower frequencies arises during matter-domination. Interestingly, for lower breaking scales future detectors tend to be most sensitive to this second, often neglected, contribution. 
For comparison we show current sensitivity from gravitional wave experiments from NANOGRAV~\cite{Arzoumanian:2018saf} and Laser Interferometer Gravitational-Wave Observatory (LIGO)~\cite{LIGOScientific:2019vic} as well as projected sensitivity from planned gravitational wave searches using the Square Kilometer Array pulsar set~\cite{Janssen:2014dka}\footnote{Supermassive black hole (SMBH) mergers may make it challenging to detect a stochastic background at the frequencies relevant for pulsar timing arrays~\cite{McWilliams_2014,2018MNRAS.473.3410R}. However, these have large uncertainties in the merger rate arising from the stellar mass function, the fraction of galaxy mergers that result in SMBH mergers, and the last parsec problem. Furthermore, since the shape of the gravitational wave spectrum of SMBH mergers ($ \Omega _{ {\rm GW}} \propto f ^{ 2/3 } $) is distinct from that of cosmic strings one could in principle attempt to disentangle the two. We assume searches are background-free in setting our constraints though note that, once gravitational waves from supermassive black hole mergers are observed, this could constitute an important background.}, Laser Interferometer Space Antenna~\cite{Caprini_2016}, Big Bang Observer~\cite{Yagi:2011wg}, DECi-hertz Interferometer Gravitational wave Observatory~\cite{Kawamura:2011zz}, Einstein Telescope~\cite{Punturo:2010zz}, Cosmic Explorer~\cite{Evans:2016mbw}, and LIGO at its design sensivitiy~\cite{LIGOScientific:2019vic}. Note throughout we present the experimental noise sensitivity. Searches for a known signal shape (as is the case for cosmic strings) can discover signals below the background. 

\begin{table}
\bgroup
\def\arraystretch{1.5}
\begin{tabular}{l c cc c cc}
\toprule[1pt]
 &&  \multicolumn{2}{c}{$H = G_{\rm SM}$} & &  \multicolumn{2}{c}{$H = G_{\rm SM} \times {\mathbb Z}_{2}$} \\ \cmidrule[0.5pt](l){3-4}  \cmidrule[0.5pt](r){6-7} 
$ G $  & & {\bf defects} & {\bf Higgs} & & {\bf defects} &  {\bf Higgs} \\ 
$G_{\rm disc}$ & & domain wall$^{*}$ & $B-L=1$ & & domain wall$^{*}$  & $B-L=2$ \\ 
$G_{B-L}$ & & abelian string$^{*}$ & $B-L=1$& &  ${\mathbb Z}_{2}$ string$^{\dagger}$ & $B-L=2$\\ 
$G_{LR}$ & & texture$^{*}$ & $({\bf 1}, {\bf 1}, {\bf 2}, \frac{1}{2})$ &&  ${\mathbb Z}_{2}$ string & $({\bf 1}, {\bf 1}, {\bf 3}, 1)$ \\ 
$G_{421}$ & & none & $({\bf 10}, {\bf 1}, 2)$ & & ${\mathbb Z}_{2}$ string & $({\bf 15}, {\bf 1}, 2)$\\ 
$G_{\rm flip}$ & & none & $({\bf 10}, 1)$ & &  ${\mathbb Z}_{2}$ string & $({\bf 50}, 2)$ \\ 
\bottomrule[1pt]
\end{tabular}
\egroup
\caption{Extended gauge symmetry and topological defects for different symmetry breaking patterns, $ G \rightarrow H $.  Whether the matter parity ${\mathbb Z}_{2}$ remains unbroken depends on the choice of the Higgs representations, and here we show examples for each case.  The defects with asterisks $*$ are unstable against tunneling effects if $G$ is embedded into a semi-simple group such as $SO(10)$ or Pati-Salam $G_{PS}$.  The ${\mathbb Z}_{2}$ string with a dagger $\dagger$ is an abelian string whose ${\mathbb Z}_{2}$ string is stable even with the embedding.  See the body of the Letter for more details.}\label{tabl:extended}
\end{table}

The projections shown here would test all breaking patterns given in Table~\ref{tabl:extended} that predict cosmic strings. In computing the spectrum we employed the approximation that $ \mu \sim v ^2 $ however for a particular symmetry breaking pattern this would change by an $ {\cal O} (1) $ factor and hence would shift the curves in Fig.~\ref{fig:stringGW} by this same $ {\cal O} ( 1 ) $ factor up/down. Nevertheless, since $ v \gtrsim  10 ^{ 10 } ~{\rm GeV} $ can be firmly tested by future experiments, such missions can probe almost the entire range relevant for thermal leptogenesis.

In principle, one could learn about the specific dynamics of leptogenesis using the cosmic string network. If leptogenesis takes place in the weak washout regime, the right handed neutrinos may dominate the energy density of the universe inducing an early period of matter domination which would be imprinted onto the GW spectrum~\cite{Cui:2018rwi}. Furthermore, they would dump entropy into the SM, diluting the present energy density of strings at the time of decay. While intriguing, in order for this to be observable with currently proposed detectors would require this period to last until temperatures of order the electroweak scale, outside of typical parameters required for leptogenesis and we do not consider it further here. 
\section{Unstable Defects}
\label{sec:instability}

When $G_{B-L}$ is embedded into simply-connected groups such as $SO(10)$ or $G_{\rm PS}$, and is broken to $G_{\rm SM}$ without the matter parity, there cannot be a stable string.  The strings are not stable against pair creation of a monopole and anti-monopole that can cut a string into two halves~\cite{Vilenkin:1982hm}. This is a tunneling process and is suppressed when the string symmetry breaking scale, $ v $ is parametrically lower than the unification scale, $  V $. Once the string is cut, the string tension quickly pulls monopoles at the two ends together forcing them to annihilate. However, this process is exponentially suppressed and if the string network is sufficiently long-lived we can expect gravitational waves.

The tunneling rate can be estimated semi-classically resulting in a rate of breaking per unit length~\cite{Monin:2008mp},
\begin{equation} 
\frac{\Gamma}{L} = \frac{\mu}{2\pi} \frac{g}{4\pi} e^{-\pi m^{2}/\mu}\,,
\end{equation} 
where $m$ is the mass of the monopole and $ g $ denotes the gauge coupling.Here we attempt only an order of magnitude estimate.  The mass of a 't Hooft--Polyakov magnetic monopole \cite{tHooft:1974kcl,Polyakov:1974ek} for $SO(3)/SO(2)$ is $m=4\pi V / g$
in the BPS limit \cite{Bogomolny:1975de,Prasad:1975kr}, and larger by an $O(1)$ constant otherwise.  On the other hand, for an abelian string in the BPS limit, both the gauge boson and Higgs mass are $e v$ and the string tension is $\mu  = \frac{4}{3}\pi v^{2}$ (see, {\it e.g.}\/, \cite{Baker:1999xn}).  For realistic groups there are $O(1)$ group theory factors which we ignore.  We also ignore the running of the gauge coupling constant between two scales.
The string network survives down to the Hubble rate
\begin{equation}
	H \sim \frac{\Gamma}{L} \ell  \sim v^{2} \ell e ^{ -12 \pi^2  V ^2 / g^{2} v^2 } \ .
\end{equation}
We make an assumption that a typical length of a string is of the Hubble size $\ell \sim H^{-1}$.  This gives,
\begin{equation}
	H \sim v e^{-6\pi ^2 V^{2} /  g^{2} v^{2}} \ .
\end{equation}
In principle, this could provide a lower cutoff on frequencies today to the frequency spectrum of GW (see, {\it e.g.}, Fig. 7 of ~\cite{Leblond:2009fq}) and provide additional emission from bursts when the string self destructs~\cite{Leblond:2009fq}. However, we see that even for a small separation between $ V $ and $ v $, there is a large exponential suppression in the rate and we can neglect this process.  Therefore the string network is expected to survive, giving us the stochastic gravitational wave signal discussed in the previous section.

Similarly, if $G_{\rm disc}$ is embedded inside a continuous group, then domain walls will be unstable against the creation of a string.  In this case, the observation of radiation domination at Big-Bang Nucleosynthesis requires the tunneling process to be fast enough to destroy all the domain walls by temperatures of order an MeV. We leave this interesting case for future study.

\section{Additional Sources}
In addition to the cosmic string network, there are also potential contributions to the stochastic gravitational waves from texture and first-order phase transitions. It is well known that textures can arise from breaking of a global symmetry \cite{Giblin:2011yh}. The produced gravitational wave spectrum is scale invariant and the peak amplitude is controlled by the seesaw scale, $v  $. Thus textures provide a unique probe of high scale physics. Furthermore, the breaking of a local symmetry can also lead to a gravitational wave spectrum arising from gauged textures~\cite{HiramatsuKohri}. For local textures, the gravitational wave spectrum is not scale invariant because the gauge field configuration cancels the gradients of the scalar field on large scales. The spectrum then has a cutoff of 
\begin{equation} 
f _0  \sim g v \frac{ a }{ a _0 } \sim 10 ^{ 11} {\rm Hz}\,,
\end{equation} 
independent of $ v $. In the absence of a higher frequency probe of gravitational waves,\footnote{Building higher frequency detectors with ability to probe physically relevant energy densities in GW is challenging since for the fixed energy density of a stochastic background, the induced characteristic strain scales inversely with the frequency.} local textures do not provide a useful test of the seesaw paradigm. 

For a first order phase transition, the gravitational wave spectrum obeys a broken power law, such that detectors are only sensitive to the spectrum near the peak frequency. The peak frequency is controlled by the temperature at the end of the transition and the inverse transition time. Assuming modest super-cooling, the acoustic source has a peak frequency \cite{Weir:2017wfa}
\begin{equation}
	f_{\rm peak} \sim  0.5~{\rm Hz} \frac{T_*}{10^{4} {\rm  GeV}}\ ,
\end{equation}
where $T_*$ is the temperature at the end of nucleation. For the seesaw scale, the range of peak frequencies predicted by a high scale $ B - L $ phase transition is much bigger than what any currently-planned gravitational wave observatories will cover. We note that in principle, a highly super-cooled transition can have a peak frequency several orders of magnitude lower due to two different effects. Firstly, super-cooling increases the duration of the phase transition leading to larger bubbles whose collisions emit lower frequency gravitational waves and secondly in such a scenario the temperature the phase transition takes places is significantly lower than the breaking scale, $T_*\ll v$. In principle, if $T_* \lesssim v /10$ high frequency gravitational wave experiments are sensitive to the lower range of parameter space $ v \sim 10^{9}$ GeV. Some work has done in this direction \cite{Jinno:2016knw,Marzo:2018nov,Brdar:2018num,Okada:2018xdh,Hasegawa:2019amx}, however, for a more generic probe of phase transitions from the seesaw scale, high-frequency gravitational wave detectors are required. Such a detector provides a unique tool to uncover physics at very early Universe, and hence should be pursued. 

\section{Conclusion and Outlook}
Thermal leptogenesis through the type I seesaw mechanism gives an elegant and minimal explanation for two outstanding puzzles in the standard model. Unfortunately, the scale of physics is naturally well beyond what we can directly test on Earth. Given the fundamental nature of these puzzles, indirect tests of thermal leptogenesis are of great value and we propose cosmic strings as a powerful probe of the paradigm. We find that, if present, gravitational wave radiation from cosmic strings can probe all the parameter space relevant for thermal leptogenesis and complements direct probes~\cite{Chun:2017spz}. Our argument is based on the simple observation that the right handed neutrino mass necessary to explain the observed neutrino masses is below the Planck or a possible grand unification scale. This suggests that some symmetry survives below to these scales to protect the right handed neutrino mass. Since successful leptogenesis requires the breaking of this symmetry to be below the scale of inflation, its breaking can be observed through its predicted cosmological defects. We show that cosmic strings often appear through the breaking of this symmetry predicting a spectrum of stochastic gravitational waves and, given our best estimates of the GW signal, future detectors are expected to probe the entire mass range relevant to the paradigm of thermal leptogenesis. While uncertainties in the gravitational wave spectrum produced by cosmic strings persist, settling this theoretical uncertainty will make GW detectors a robust probe of thermal leptogenesis.

Once the spectrum of the stochastic gravitational wave is mapped out, we should be able to remove any contributions from astrophysical sources peaked at specific frequencies. A cosmic string network predicts a nearly scale-invariant spectrum over many decades and would be clear indication of a symmetry broken at a high scale, with the amplitude and cutoff corresponding to the symmetry breaking scale. If such a spectrum is discovered and falls into the energy scales relevant for the seesaw mechanism and leptogenesis, it would provide intriguing hints of dynamics in the lepton sector at high scales.

\begin{acknowledgments}

We thank Wilfried Buchm\"uller, Valerie Domcke, G\'eraldine
Servant, Ryusuke Jinno, Yann Gouttenoire, Kohei Kamada, and Djuna Croon for useful
discussions.  This work was supported by the NSF grant PHY-1638509
(H.M.), by the U.S. DOE Contract DE-AC02-05CH11231 (H.M.), by the JSPS
Grant-in-Aid for Scientific Research JP17K05409 (H.M.) and JP17H01131
(K.K.), MEXT Grant-in-Aid for Scientific Research on Innovative Areas
JP15H05887 (H.M.), JP15K21733 (H.M.), JP15H05889 (K.K.), JP18H04594
(K.K.), JP19H05114 (K.K.), by WPI, MEXT, Japan (H.M., and K.K.), Hamamatsu Photonics (H.M.), DE-AC02-05CH11231 (J.D.) and TRIUMF (G.W.), which receives federal funding via the NRC.
\end{acknowledgments}

\bibliographystyle{utcaps_mod}
\bibliography{seesaw}

\end{document}